\documentclass{PoS}

\usepackage[utf8]{inputenc}
\usepackage[T1]{fontenc}

\title{Status and discovery prospects for light
pseudoscalars in the NMSSM}

\ShortTitle{Light pseudoscalars}

\author{ R.~Aggleton\\
				University of Southampton, UK\\
				Rutherford Appleton Laboratory, UK\\
				Bristol university, UK\\
        E-mail: \email{robin.aggleton@cern.ch}}

\author{ D.~Barducci\\
        SISSA and INFN, Sezione di Trieste, via Bonomea 265, 34136 Trieste, Italy\\
        E-mail: \email{daniele.barducci@lapth.cnrs.fr}}

\author{\speaker{N.-E. Bomark}\\
        University of Agder, Norway\\
        E-mail: \email{nilseb@uia.no}}

\author{ S.~Moretti\\
				University of Southampton, UK\\
        E-mail: \email{S.Moretti@soton.ac.uk}}

\author{ C.~Shepherd-Themistocleous\\
				University of Southampton, UK\\
				Rutherford Appleton Laboratory, UK\\
        E-mail: \email{Claire.Shepherd@stfc.ac.uk}}

\abstract{While most BSM searches at the LHC focus on heavy new states, the NMSSM contains the possibility of new light states that have escaped detection due to their singlet nature.\\
Here we focus on light pseudoscalars, investigating the parameter space impact of recent LHC searches for such light states stemming from the decay of the 125 GeV Higgs boson. It is shown that, though direct searches can not yet compete with the requirement of the 125 GeV scalar having SM-like couplings, the searches are touching the allowed parameter space and should make a phenomenological impact in the near future.}

\FullConference{EPS-HEP 2017, European Physical Society conference on High Energy Physics\\
		5-12 July 2017\\
		Venice, Italy}

\begin{document}

\section{Introduction}
Five years after the discovery of the Higgs-boson, we have seen nothing but Standard Model (SM) physics from the LHC. So should we expect anything more, and if so, what?

The only problem we have that is really pointing towards new physics at the electroweak (EW) scale, is the hierarchy problem; other problems, like dark matter, may have its solution at this scale too, but nothing points specifically towards something discoverable at the LHC\footnote{But what about the WIMP miracle you may ask, does it not point to EW scale dark matter?\\ Maybe, but there are many other perfectly fine possibilities for dark matter and WIMPs may not necessarily show up at the LHC either. Also the popularity of the WIMP is related to the expectation of new physics at the EW scale due to the hierarchy problem.}. As far as solutions to the hierarchy problem are concerned, Supersymmetry (SUSY) sticks out; the absence of sparticles at the LHC is becoming more and more of a problem, but one should remember that all other theories for new physics have the exact same problem, and hence it is fair to say that SUSY is still our best bet for the future, although it is getting more likely that nothing new will show up and we have misunderstood something fundamental.

One necessary ingredient in the superpotential of SUSY models, is a $\mu$ term, $\mu \widehat{H}_u\widehat{H}_d$, where $\mu$ is a dimensional parameter that needs to be at the EW scale in order for the EW symmetry breaking to work. Although $\mu$ is far from the only dimensional parameter in SUSY theories, it is special as it respects SUSY; all other mass parameters are soft SUSY breaking terms.

This poses a problem, why should $\mu$, that has nothing to do with SUSY breaking, have the same scale as the SUSY breaking terms?

One solution is offered by the Next to Minimal SUSY SM (NMSSM)~\cite{Ellwanger:2009dp}; forbid the $\mu \widehat{H}_u\widehat{H}_d$ term and instead introduce a singlet complex scalar field $S$ and the term, $\lambda \widehat{S}\widehat{H}_u\widehat{H}_d$. By giving $S$ a VEV --- which comes from soft SUSY breaking terms and is hence naturally at the EW scale --- we then get an effective $\mu$ term, $\lambda \langle S\rangle{H}_u{H}_d$.

The additional singlet superfield introduces two terms into the superpotential,
\begin{equation}\label{eq:SuperPot}
  W_{\mathrm{NMSSM}} \supset \lambda \widehat{S}\widehat{H}_u\widehat{H}_d + \frac{\kappa}{3}\widehat S^3,
\end{equation}
where $\lambda$ and $\kappa$ are dimensionless coupling constants.

In addition the soft SUSY breaking potential needs to be supplemented with,
\begin{equation}\label{eq:SoftHiggs}
V^\mathrm{NMSSM}_{\mathrm{soft}} \supset m_{S}^2 | S |^2 + \left( \lambda A_\lambda H_u H_d S + \frac{\kappa}{3} A_{\kappa}S^3 +  \mathrm{h.c.}\right)\; , \\
\end{equation}
where $m_S$, $A_\lambda$ and $A_\kappa$ are dimensionful mass and trilinear parameters. The singlet mass term, $m_{S}^2 | S |^2$ is traded for an effective $\mu$ term and hence the set of parameters relevant for the scalar sector used in the following scans are, $\lambda,\kappa,\mu_{\rm eff}, \tan\beta, A_\lambda$ and $A_\kappa$, where $A_\kappa$ in some scans have been replaced by the diagonal entry of the pseudoscalar mass matrix, $M_p$, more details on the scan procedures can be found in~\cite{Aggleton:2016tdd}.

\section{Light pseudoscalars}

Even though many parameters do affect the pseudoscalar mass, it is mostly driven by $A_\kappa$. Since $A_\kappa$ is basically unconstrained by other observables, we can put it close to zero in order to get a light pseudoscalar. This means that in most of parameter space, a light pseudoscalar is easily achieved.

Such light pseudoscalars are always very singlet-like and hence hard to produce, 
which is why they may have escaped detection so far. The only direct production 
channel of interest is associated production, $b\bar b a_1$ with $a_1$ being the lightest pseudoscalar, but whether this
channel is usable is still an open question~\cite{Bomark:2014gya}.

The most promising searches look for pseudoscalars in the decay of other 
particles, especially heavier scalars. Which scalar one should best assume to 
start such a chain is not clear, but since the only one actually known to exist 
is the $h_{125}$, that is a good place to start.

In the following we will therefore look at the channel $gg\to h_{125}\to a_1a_1$.

\section{Limits on Br($h_{125}\to a_1a_1$)}

The most important value for this channel is Br($h_{125}\to a_1a_1$), the 
production cross section $\sigma(gg\to h_{125})$ does not change much 
throughout the parameter space and the branching ratios for the decay channels 
of the pseudoscalars are pretty much fixed once the mass is known. Br($h_{125}\to a_1a_1$) on the other hand, can vary all the way from 0 to 1.

Since a large Br($h_{125}\to a_1a_1$) will suppress the other branching 
ratios of $h_{125}$, the most important experimental constraints here are 
the signal rate constraints for the 125 GeV Higgs. In NMSSMTools~\cite{Ellwanger:2005dv,Belanger:2005kh} these are 
implemented as three separate constraints on the $ZZ$, $\gamma\gamma$ and $b\bar 
b$ reduce couplings, taken from Lilith~\cite{Bernon:2015hsa}. HiggsSignals~\cite{Bechtle:2013xfa} on the other hand, does an 
overall fit to all channels simultaneously.

The result is that NMSSSMTools allows Br($h_{125}\to a_1a_1)<0.2$ while HiggsSignals allows Br($h_{125}\to a_1a_1)<0.5$!

Which of these values is most trustworthy is hard to say, so we leave that as is and show both options when comparing to data.

\begin{figure}
    \centering
        \includegraphics[width=0.47\textwidth]{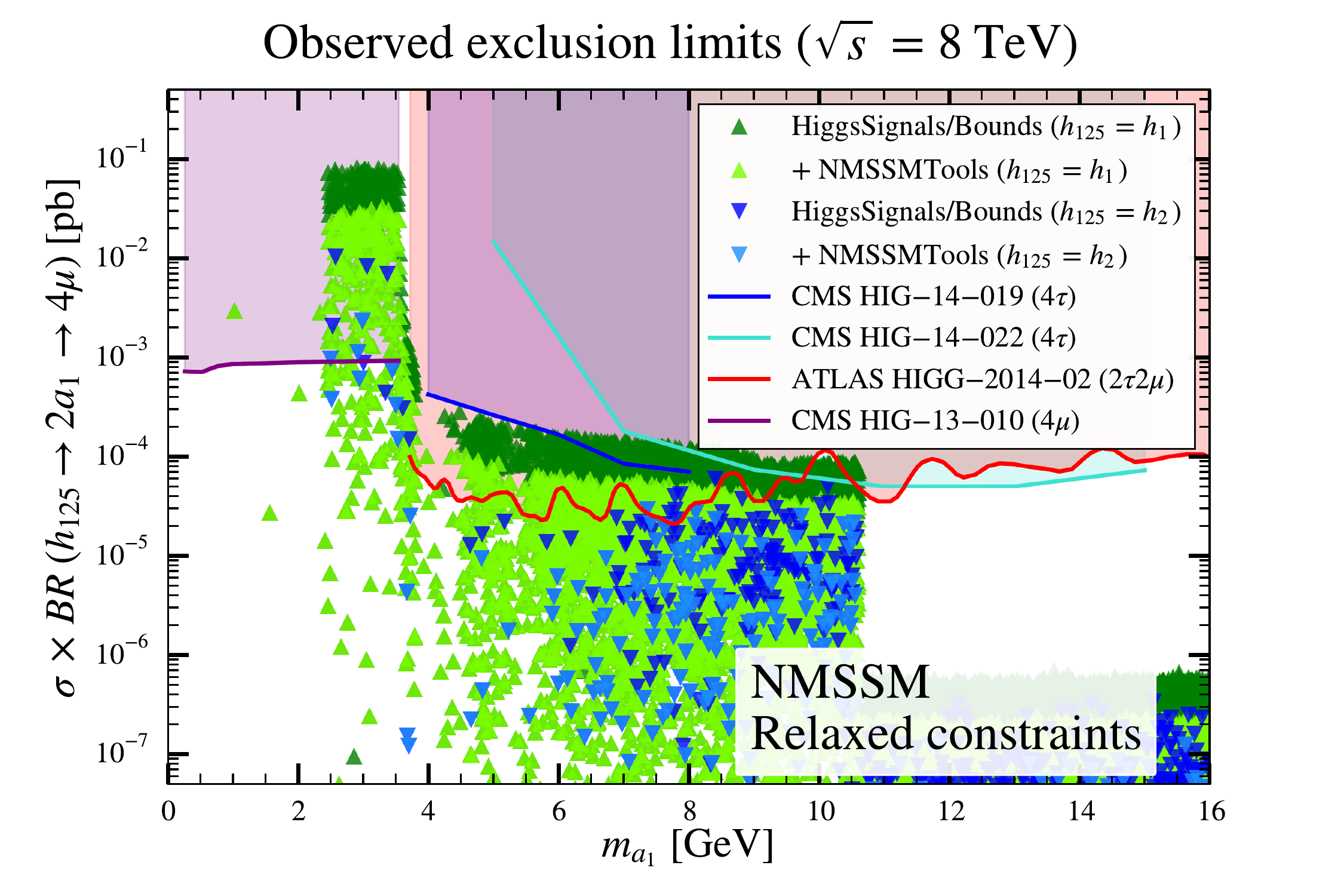}
        \includegraphics[width=0.47\textwidth]{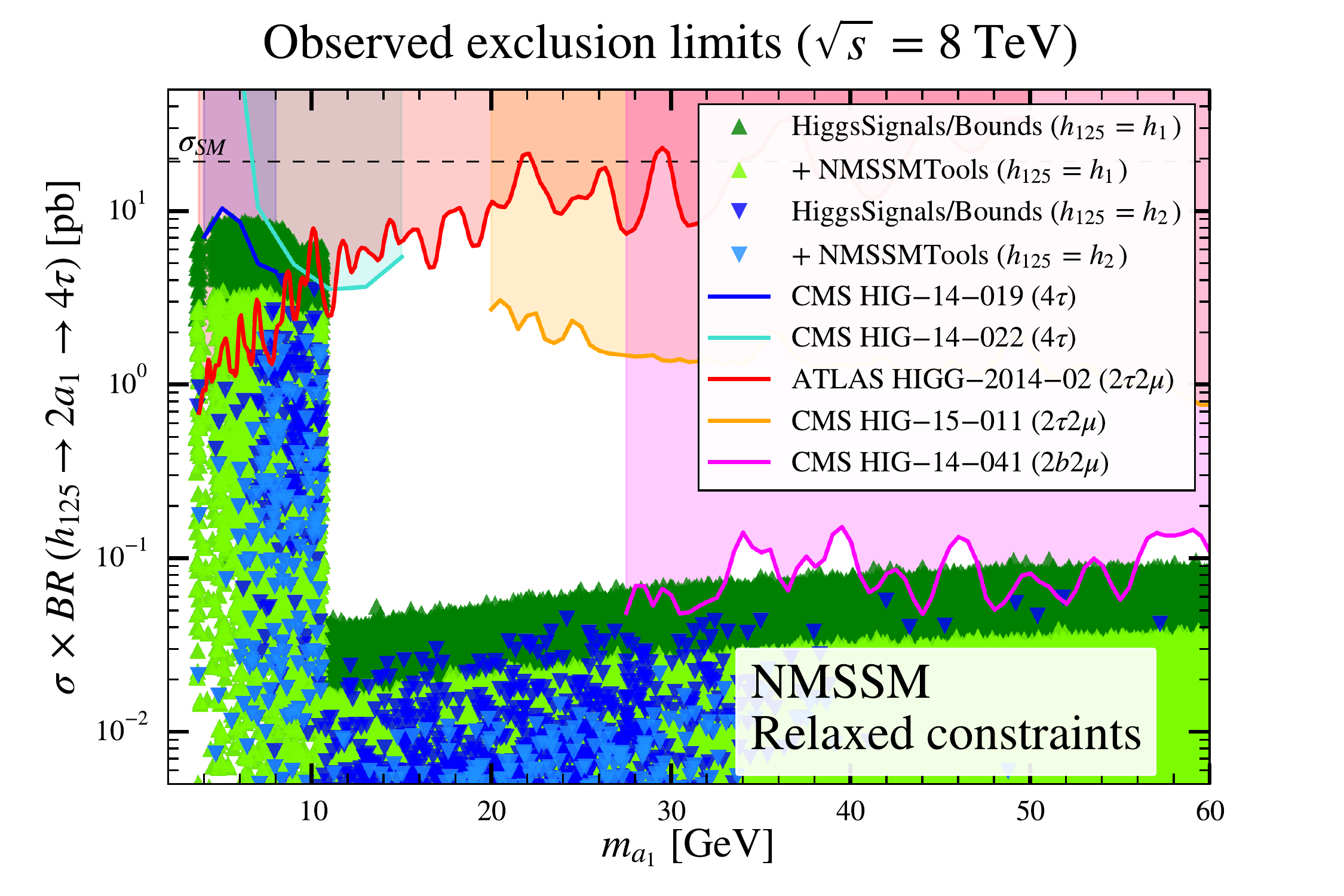}
    \caption{Plots of $\sigma \times BR(gg \to h_{125} \to 2a_1 \to 4\mu)$ (left) and $\sigma \times BR(gg \to h_{125} \to 2a_1 \to 4\tau)$ (right) versus $m_{a_1}$ for various Higgs assignments in the NMSSM. Dark green/blue points are only required to satisfy Higgs rate constraints from HiggsSignals, whilst lighter green/blue points must also pass NMSSMTools Higgs rate constraints. All points pass a ``relaxed'' set of constraints, i.e.\ all other NMSSMTools constraints, but ignoring $(g-2)_\mu$ and only an upper limit on relic density. Overlaid are observed exclusion regions from the relevant analyses. }
     \label{fig:Limits}
\end{figure}

\section{Experimental limits}

Which final state to look for depends on the pseudoscalar mass. Below $2m_\tau$ looking for muons is rather promising. In the left panel of figure~\ref{fig:Limits}, we can see that the $4\mu$ searches~\cite{Khachatryan:2015wka} puts serious pressure on the parameter space. 

If we move to the range $2m_\tau<m_{a_1}<2m_b$, one can still exploit the detectability of muons as well as taus. As can be seen in figure~\ref{fig:Limits}, searches for $2\tau2\mu$~\cite{Aad:2015oqa} are starting to cut into the parameter space, though the impact is not that large yet.

In the higher end, $2m_b<m_{a_1}$, the presence of $b$-quarks in the final state makes detection harder, but from the right panel of figure~\ref{fig:Limits}, we can see that $2b2\mu$ searches~\cite{CMS-PAS-HIG-14-041} are starting to have an impact, at least if we accept Br($h_{125}\to a_1a_1$) up to 0.5 as HiggsSignals does.

All in all there is an impressive experimental effort to constrain also these low mass new states.

\section{Conclusions}
Light scalars and pseudoscalars are possible in well motivated
theories for new physics, especially in the NMSSM. This presents a different kind of challenge for the LHC experiments as compared to the standard heavy new physics.

Since such light particles would interact very weakly, they can easily have escaped detection and for the same reason will be hard to find at the LHC. One of our best options to find them, is by looking at cascade decays of heavier particles, especially looking for $h_{125}\to a_1a_1$.

While these searches are difficult because of the soft final states and the fact that pseudoscalars above 10 GeV mostly decays to $b$-quarks, both ATLAS and CMS are making progress towards constraining this parameter space and some of the low mass searches are already excluding regions of parameter space.


\begin{thebibliography}{99}

\bibitem{Ellwanger:2009dp}
U.~Ellwanger, C.~Hugonie, and A.~M. Teixeira, 
	{\em Phys. Rept.} {\bf 496} (2010) 1--77,
  [\href{http://xxx.lanl.gov/abs/0910.1785}{{\tt 0910.1785}}].

\bibitem{Aggleton:2016tdd}
  R.~Aggleton, D.~Barducci, N.~E.~Bomark, S.~Moretti and C.~Shepherd-Themistocleous,
  {\em JHEP} {\bf 02} (2017) 035,
  [\href{http://xxx.lanl.gov/abs/0910.1785}{{\tt 1609.06089}}].

\bibitem{Bomark:2014gya}
N.-E. Bomark, S.~Moretti, S.~Munir, and L.~Roszkowski, 
	{\em JHEP} {\bf 02} (2015) 044,
  [\href{http://xxx.lanl.gov/abs/1409.8393}{{\tt 1409.8393}}].



\bibitem{Ellwanger:2005dv}
U.~Ellwanger and C.~Hugonie, 
  {\em Comput. Phys. Commun.} {\bf 175} (2006) 290--303,
  [\href{http://xxx.lanl.gov/abs/hep-ph/0508022}{{\tt hep-ph/0508022}}].

\bibitem{Belanger:2005kh}
G.~Belanger, F.~Boudjema, C.~Hugonie, A.~Pukhov, and A.~Semenov, 
	{\em JCAP} {\bf 0509} (2005) 001,
  [\href{http://xxx.lanl.gov/abs/hep-ph/0505142}{{\tt hep-ph/0505142}}].

\bibitem{Bernon:2015hsa}
J.~Bernon and B.~Dumont, 
	{\em Eur. Phys. J.} {\bf C75} (2015), no.~9 440,
  [\href{http://xxx.lanl.gov/abs/1502.04138}{{\tt 1502.04138}}].


\bibitem{Bechtle:2013xfa}
P.~Bechtle, S.~Heinemeyer, O.~Stål, T.~Stefaniak, and G.~Weiglein, 
	{\em Eur. Phys. J.} {\bf C74} (2014), no.~2 2711,
  [\href{http://xxx.lanl.gov/abs/1305.1933}{{\tt 1305.1933}}].
	
\bibitem{Khachatryan:2015wka}
{\bf CMS} Collaboration, V.~Khachatryan {\em et.~al.}, 
	{\em Phys. Lett.} {\bf B752} (2016) 146--168, [\href{http://xxx.lanl.gov/abs/1506.00424}{{\tt  1506.00424}}].
	
\bibitem{Aad:2015oqa}
{\bf ATLAS} Collaboration, G.~Aad {\em et.~al.}, 
	{\em Phys. Rev.} {\bf D92}  (2015), no.~5 052002, [\href{http://xxx.lanl.gov/abs/1505.01609}{{\tt  1505.01609}}].

	
\bibitem{CMS-PAS-HIG-14-041}
{\bf CMS} Collaboration, 
  Tech. Rep. CMS-PAS-HIG-14-041, CERN, Geneva, 2016.

	
\end{thebibliography}
\end{document}